\documentstyle[11pt,epsfig,psfig]{article}
\newcommand{\AmS}{{\protect\the\textfont2  A\kern-.1667em\lower.5ex\hbox{M}\kern-.125emS}}
\begin{document}
\title{\huge{Prospects for probing the gluon density in protons 
using heavy quarkonium hadroproduction}
\thanks{Research partially supported by CICYT under grant AEN99-0692}} 
\author{{\bf J.L. Domenech-Garret$^{a}$\thanks{domenech@evalo1.ific.uv.es}
$\ $, M.A. Sanchis-Lozano$^{b,c}$\thanks{mas@evalo1.ific.uv.es}$\ $ and
S. Wolf $^c$\thanks{Stefan.Wolf@ific.uv.es}}
\\ \\
\it (a) Departamento de F\'{\i}sica At\'omica, Molecular y Nuclear \\
\it (b) Instituto de F\'{\i}sica
 Corpuscular (IFIC), Centro Mixto Universidad de Valencia-CSIC \\
\it (c) Departamento de F\'{\i}sica Te\'orica \\
\it Dr. Moliner 50, E-46100 Burjassot, Valencia (Spain) }
\date{}
\maketitle 
\begin{abstract}
        We examine carefully bottomonia hadroproduction in
        colliders as a way of probing the gluon density in
        protons. To this end we develop some previous work,
        getting quantitative predictions and concluding that 
        our proposal can be useful to perform consistency checks
        of the parameterization sets of different parton distribution 
        functions.
\end{abstract}
\vspace{-12.5cm}
\large{
\begin{flushright}
  IFIC/01-54\\
  FTUV-01-1005\\
  February 12, 2002
\end{flushright} }
\vspace{14cm}
\newpage
\section{Introduction}

With the advent of large hadron colliders, 
a precise knowledge of the structure of hadrons has become
compulsory if experimental data on elementary particle
physics are expected to be analysed and ultimately
interpreted with the desired high
accuracy and reliability. In particular, 
one of the goals of the LHC project is to perform precise tests of the
Standard Model of strong, weak and electromagnetic interactions and
the fundamental constituents of matter. Actually  
the LHC machine can be viewed as a gluon-gluon collider to a
large extent and
many signatures (and their backgrounds) of physics, both within and 
beyond the Standard Model, involve gluons in the initial state
\cite{tdr}. Therefore an accurate 
knowledge of the gluon density in protons acquires a
special relevance. Currently there are three 
major groups - namely CTEQ, MRST and GRV - providing
regular updates of the partonic structure of protons
as new data and/or theoretical improvements
become available \cite{thorne}.

The parton distribution functions
(PDFs) are phenomenologically determined
by global analyses of a wide class of hard processes
involving initial-state hadrons, making use of the
QCD parton framework.  
Presently, the most precise determinations of the gluon momentum
distributions in the proton basically
come from data on
deep-inelastic scattering (DIS), in particular through the
analysis of the
scaling violations of the structure function $F_2$. However, 
this represents an indirect method since the gluon density is obtained
by means of the QCD evolution (DGLAP) equations. On the other hand,  
hadron-hadron scattering processes with prompt photon production
or jets in the final state should be (and actually already are)
extremely adequate to
probe $\lq\lq$directly'' the gluon distribution in hadrons 
\cite{babuk}.  
Such $\lq\lq$direct'' determinations can be viewed as complementary
to indirect analyses, providing independent tests  of perturbative
QCD with distinct systematic errors. Let us also
remark that, in colliders like the Fermilab Tevatron or the 
LHC \cite{tapprogge}, gluon densities are - or can be - probed at similar
(and higher) $x$-values as in DIS but at significantly larger 
energy scales.

In this paper, we examine the possibility of using heavy quarkonia 
inclusive hadroproduction to probe the gluon density of protons, extending
in a more quantitative way the ideas earlier
presented in Refs.\cite{mas6,mas01}. In particular we are focusing on
bottomonia production in proton-proton collisions at the LHC.
Our claim is that, ultimately, the measurement of bottomonia
cross sections could provide a useful consistency check
of different PDF sets and their energy-scale evolution.

\section{Heavy quarkonia inclusive hadroproduction}

At very high transverse momentum, fragmentation mechanisms should
expectedly become dominant in the cross section of bottomonium 
hadroproduction - as in any other single-particle production channel. 
Specifically, gluon fragmentation into a $b\bar{b}$ pair followed by
its non-perturbative evolution yielding a $\Upsilon$ final state by
emission of soft gluons, should 
play the leading role according to the colour-octet mechanism (COM)
\cite{braaten}. Such a production mechanism constitutes
the relativistic generalization of the colour-singlet model (CSM)
and one of the most natural explanations of the excess of
heavy quarkonia production observed at the Tevatron \cite{fermi,cho,mas97}. 
Moreover, the COM can be cast into a rigorous framework based on
NRQCD \cite{bodwin}, an effective theory coming from first principles.

Furthermore, as long as we focus on quarkonium hadroproduction 
at large transverse
momentum, the factorization assumption, underlying the NRQCD description, 
should be justified by the large scale $p_T$; for bottomonia
in particular the heavy quark mass is likely large enough to 
factorize the short and long distance physics in the partonic
interaction itself, as it is well known from
decay processes. Whether similar arguments can be applied to 
charmonium resonances has to
be checked, for example analysing the transverse polarization of
the resonance \cite{beneke2}.

Therefore one can expect, on solid grounds, that one of the
 main production channels of bottomonia at the LHC
should correspond to the partonic subprocess:
\begin{equation}
g\ g \ {\rightarrow}\ g^{\ast}\ g
\end{equation}
and the subsequent gluon fragmentation into a $\Upsilon(nS)$ state
accompanied by light hadrons $X_s$:
\begin{equation}
 g^{\ast}{\rightarrow}\ \Upsilon(nS)\ X_s\ \ \ \ \ ;\ \ \ \ (n=1,2,3) 
\end{equation}
produced through the already mentioned colour-octet mechanism. 

We must clearly state that our later
development relies on the hypothesis that the main
contribution to heavy quarkonium hadroproduction at high
transverse momentum comes from a fragmentation
mechanism as in (2) whose final hadronization stage is governed by
a single non-perturbative parameter. 
Strictly speaking, the colour-octet production mechanism
is not necessarily required; for example, another possibility is gluon
fragmentation into a heavy resonance via
a colour-singlet channel. However, this mechanism actually 
corresponds to a higher order contribution
since colour conservation and charge conjugation require the 
emission of two extra
gluons in the subprocess (2) \cite{kraemer}. 
Therefore, the colour-octet fragmentation mechanism 
should play the leading role at high transverse
momentum in bottomonium hadroproduction \cite{mas01} if
the COM turns out to be correct. In this regard, the different
leading and subleading contributions and their relative importance 
will be very briefly reviewed in section 2.1.

Ideally, the final-state gluon ($g$) in Eq.(1)
will give rise to a recoiling jet ($g\ {\rightarrow}\ jet$), 
sharing, in principle, the same 
transverse momentum as the heavy resonance - neglecting 
initial and final state gluon radiation. On the other hand, the
leptonic decay of the resonance is obviously the best way
to have a clean signal of its formation among a huge
hadronic background. Of particular interest is the muonic
channel, since such muons would very likely 
pass the first level trigger - consisting (in ATLAS) of a muon with
transverse momentum larger than 6 GeV and (absolute)
pseudo-rapidity less than 2.5 \cite{tdr}.

Hence, events would topologically consist of an almost
isolated muon pair from the decay of the heavy resonance recoiling 
against a single jet, therby providing a suitable tag for the production 
mechanism represented in Eqs.(1-2). Indeed
this topology is the expected one since the
mass difference between the intermediate coloured and
final states should be quite small, allowing the emission
of at most a few light hadrons via soft gluon radiation - denoted
by $X_s$ - accompanying the resonance at the
final hadronization stage. Moreover, in order to remove possible
background, events with (more than one) mini-jets should be
discarded from the selected sample.
\vskip 1.cm

\begin{figure}[htb]
\centerline{\hbox{
\psfig{figure=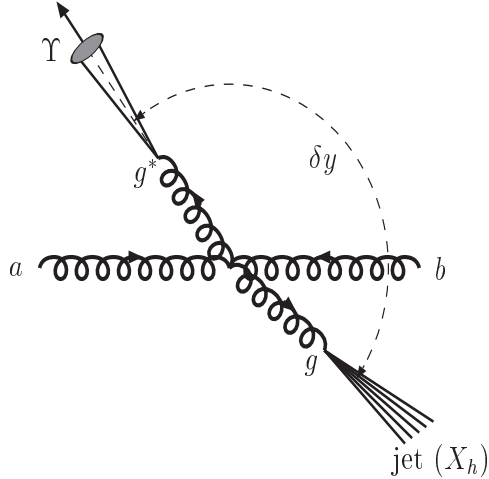,height=6.5cm,width=6.5cm}
}}
\caption{Sketch of a parton-level process leading to
a resonance recoiling against a gluon yielding a final-state jet. 
Notice that if the two interacting partons had different 
fractions $x$ of the total momenta of the colliding protons, the
event would not show a back-to-back topology between the dimuon and
the recoiling jet. Besides, initial and final state radiation
can affect the event topology similarly. We consider an
event compatible with a back-to-back topology if
the rapidity difference in absolute value of the 
final-state resonance and jet is
less than a predetermined value set by $2y_{cut}$.} 
\end{figure}

\subsection{Differential cross section}

In this Section we reproduce the main points presented in
Ref.~\cite{mas01} for the sake of clarity in our later
more technical proposal and critical discussion.

According to the collinear approximation, employed in our analysis, the 
triple differential cross 
section for the inclusive production process $pp\ {\rightarrow}\ {\Upsilon}X$ 
can be written as
\begin{equation}
\frac{d^3\sigma}{dy_{\Upsilon}dy_{jet}dp_T}= 2p_T
\sum_{ab}x_ax_bf_{a/p}(x_a,Q^2)f_{b/p}(x_b,Q^2) 
\frac{d\hat{\sigma}_{ab}(Q^2)}{d\hat{t}}
\end{equation}
where $p_T$ denotes the transverse momentum of both the
resonance and the jet; $y_{\Upsilon}$ and $y_{jet}$
represent the corresponding rapidities in the Lab frame, respectively;
$f_{a,b/p}(x_{a,b},Q^2)$ denotes the parton-$a,b$ density in the proton
at Feynman $x_{a,b}$ and typical hard interaction scale $Q^2$.
(Let us note that all factorization, fragmentation
and renormalization scales have been
taken equal for simplicity.)

The  partonic differential cross section can be written as:
\begin{equation}
\frac{d\hat{\sigma}_{ab}(Q^2)}{d\hat{t}}\ {\equiv}\ 
\frac{d\hat{\sigma}}{d\hat{t}}
(ab{\rightarrow}{\Upsilon}X_h)=\frac{1}{16{\pi}\hat{s}^2}\ 
\sum_n\overline{\sum}{\mid}{\cal A}(ab{\rightarrow}[b\bar{b}]_nX_h){\mid}^2
\langle{\cal O}_n^{\Upsilon}\rangle
\end{equation}
consisting of a short-distance (and calculable) part, 
i.e. ${\cal A}(ab{\rightarrow}[b\bar{b}]_nX_h)$ where
$X_h$ denotes the hadron bulk from the hard interaction, as
depicted in Fig.1, and a long-distance 
piece $\langle{\cal O}_n^{\Upsilon}\rangle$ which can be identified as a 
NRQCD matrix element \cite{bodwin}. The barred summation represents an
average over initial spin and colour for the short-distance
partonic process, and there is another
summation running over channels corresponding
to different quantum numbers of the intermediate state (labelled by $n$)
evolving into final heavy quarkonium. Using the 
spectroscopic notation for the $[b\bar{b}]_n$ 
state: ${}^{2S+1}\!L_J^{(1,8)}$, where the superscript $(1,8)$
stands for either a colour-singlet or a colour-octet state, the different 
contributions are:
${}^1\!S_0^{(1)}$, ${}^3\!S_1^{(8)}$, ${}^1\!S_0^{(8)}$, 
${}^3\!P_J^{(8)}$, ...
The relative importance of each contribution is basically given by
its $p_T$ dependence, the order of the strong coupling constant $\alpha_s$
and the so-called velocity scaling rules of NRQCD \cite{bodwin}
(a nice overview can be found in \cite{kraemer}). 
In this work, we shall concentrate only on 
direct production at high-$p_T$, 
(dominated by the ${}^3\!S_1^{(8)}$ contribution)   
without feedown from $\chi_{bJ}$ states.

The basic kinematics of the partonic subprocess 
(see Figure 1) is determined by
the transverse momentum $p_T$ and the rapidities of the
resonance $y_{\Upsilon}$ and of the recoiling jet $y_{jet}$,
respectively
\footnote{According to energy/momentum leading-order balance 
of the parton-level
interaction, we assume that the rapidity of the parent gluon equals
the jet rapidity which can be obtained from
its measured pseudo-rapidity $\eta_{jet}\ {\simeq}\ y_{jet}$ 
(for technical details about the measurement of the
jet pseudorapidity in the hadronic calorimeter we refer 
the reader to \cite{tdr}).}; therefore
$y_{\Upsilon}+y_{jet}$ is twice the rapidity of the
partonic centre-of-mass system $y_0$ (i.e.~$y_0=(y_{\Upsilon}+y_{jet})/2$) 
and $\delta y=y_{\Upsilon}-y_{jet}$ denotes the rapidity difference.

At large transverse momentum one can write:
\begin{equation}
x_{a,b}\ \simeq\ \frac{2p_T\ e^{{\pm}y_0}}{\sqrt{s}}\ \cosh{\frac{\delta y}{2}}
\end{equation}
where the measured transverse momentum of the
resonance $p_T$ is assumed to
coincide at leading order with the transverse energy of the jet.
Eq.(5) can be seen as a  particular application of well-known
formulae for dijet hadroproduction (see \cite{stenzel} and
references therein).

At relatively high $p_T$ the dominant partonic subprocess for $\Upsilon$
hadroproduction should be the gluon-gluon interaction \cite{mas01}. 
Thereby, the
following combination appears in the cross section of Eq.(3) 
particularizing to the case of gluon colliding partons  
$a$ and $b$: $f_{g/p}(x,Q^2)$ ${\equiv}\ g(x,Q^2)$, 
\[ x_ax_b\ g(x_a,Q^2)\ g(x_b,Q^2) \] 
calculated at the common scale $Q^2$,  
basically set by $p_T^2$ as we shall see. 

In particular, requiring $y_0=0$ (i.e.~$|y_{\Upsilon}|=|y_{jet}|$ but 
with opposite signs) implies considering only those values 
$x\ {\equiv}\ x_a=x_b$ (within the uncertainty interval to be discussed
below) given by
\begin{equation}
x\ =\ x_T\ \cosh{y_{\Upsilon}}
\end{equation}
with $x_T=2p_T/\sqrt{s}$. Thus the theoretical and experimental analysis 
simplifies notably: only the 
transverse momentum and rapidity of the $\Upsilon$ resonance
enter into Eq.(6), both determined from the experimental measurement of
the transverse momenta and pseudo-rapidities of 
the muons from its decay. 

On the other hand, the Feynman $x$ can be known with
a precision limited by the rapidity uncertainty
of the resonance; indeed, for a fixed $p_T$ value, the  
typical uncertainty ${\Delta}x$ is given by \cite{mas01}
\begin{equation}
\frac{{\mid}{\Delta}x{\mid}}{x}\ <\ y_{cut}
\end{equation}
where $y_{cut}$ represents the half-width of the allowed 
rapidity interval 
for the resonance, corresponding to a back-to-back topology
as shown in Fig.1
\footnote{We have assumed that the uncertainty
in the determination of the Feynman $x$ value provided by Eq.(7)
determines the binning of this variable. Thus, the
$x$-bin size is ultimately set by the value of $y_{cut}$, used to
impose the back-to-back condition on the events (i.e. the 
measured rapidities of the resonance and jet should differ
in absolute value by less that twice $y_{cut}$). 
We have explicitly checked that the difference, event by event,
between the $\lq\lq$observed'' $x_{a,b}$ (through Eq.(5)) after 
applying ATLFAST jet and $\Upsilon$ reconstruction \cite{tdr}, and 
the corresponding
values selected by PYTHIA in the hard partonic interaction, is
of the order of $10\%$, compatible with our choice $y_{cut}=0.2$.
(we are indebted to F. Camarena for his valuable help in this
calculation).}.

Then we can write as a first approximation,
\begin{equation}
\frac{d^3\sigma}{dy_{\Upsilon}dy_{jet}dp_T}= 2p_T\ 
x^2g(x,Q^2)^2\ \frac{d\hat{\sigma}_{gg}(Q^2)}{d\hat{t}}
\end{equation}

On the other hand we can choose the typical hard scale as the
momentum transfer $Q^2=-\hat{t}$, 
where $\hat{t}$ stands for the Mandelstam variable of the
partonic subprocess.
This quantity can be estimated by means of the expression
\begin{equation}
Q^2\ =\  
2p_T^2\ \cosh^2\frac{\delta y}{2}\  
\biggl(1-\tanh{\frac{\delta y}{2}}\biggr)\ =\  
2p_T^2\ \cosh^2y_{\Upsilon}\ (1-\tanh{y_{\Upsilon}}) 
\end{equation}
the last step coming since $\delta y=2y_{\Upsilon}$
as we have assumed $y_0=0$.

Let us remark that $x$ and $Q^2$ are not completely independent 
since the centre-of-mass energy of the partonic interaction
$\hat{s}=x^2s$  can be expressed as $\hat{s}=4p_T^2\cosh^2y_{\Upsilon}$, 
neglecting masses \footnote{We have neglected masses only
in the estimate of $Q^2$, not in the generation of events
with PYTHIA. Actually for $p_T^2{\gg}M_{\Upsilon}^2$, this leads
to no significant difference in the former case.}.
Therefore the following relation
is satisfied
\begin{equation}
Q^2=\frac{s}{2}\ x^2\ (1-\tanh{y_{\Upsilon}})
\end{equation} 

Moreover, keeping $x$ fixed and averaging over the full rapidity range,
we get
\begin{equation}
{\langle}Q^2{\rangle}\ =\ \frac{s}{2}\ {x^2}
\end{equation}
In fact, in our later development we will integrate cross
sections over the full available rapidity range and 
this average value for $Q^2$ will be extensively used.

\section{Probing the gluon density in the proton}

In Ref.~\cite{mas01}, we proposed to test different parameterizations
of the gluon distribution $g(x,Q^2)$ in the proton
by studying the ratios
\begin{equation}
\frac{x_j^2\ g(x_j,Q_j^2)^2}{x_i^2\ g(x_i,Q_i^2)^2}\ =\  
\biggl(\frac{d\hat{\sigma}_{gg}/d\hat{t}_i}
{d\hat{\sigma}_{gg}/d\hat{t}_j}\biggr){\times}
\biggl(\frac{p_{Ti}}{p_{Tj}}\biggr){\times}
\biggl(\frac{d^3{\sigma}/dy_{\Upsilon}dy_{jet}dp_{Tj}}
{d^3{\sigma}/dy_{\Upsilon}dy_{jet}dp_{Ti}}\biggr)_{y_{jet}=-y_{\Upsilon}}
\end{equation}
corresponding to several $x_i,x_j$ pairs of the parameterization set
under consideration. Subin-dices $i,j$ denote discrete values or
points of the $x$ variable
associated to bins whose relative widths are set by Eq.(7); 
$Q_i^2$ stands for the (average) transverse scale according to
Eq.(11) for $x=x_i$.
Actually the ratio given by Eq.(12) involves the so-called
effective gluon distributions \cite{combridge}, since quark distributions can
contribute to some extent via the partonic scattering channel
$gq{\rightarrow}g^{\ast}q$, and to a much lower extent
via the channel $q\overline{q}{\rightarrow}g^{\ast}g$ \cite{mas01}.

It is important to note that the number of independent $x$ values
is basically limited by ${\Delta}x$ (i.e. by $y_{cut}$), and
hence the number of available $x_i,x_j$
pairs in the ratio (12); therefore this condition cannot be 
released too much. At the
same time, the foreseen statistics does not allow one to 
decrease $y_{cut}$ too much
either, and a compromise  should be reached (our particular choice has been
$y_{cut}=0.2$).

In sum, the keypoint of our proposal is to consider the l.h.s.~of 
Eq.(12) as an {\em input} corresponding to different sets of the
gluon distribution for the proton, whose $x$ dependence is hence
assumed to be $\lq\lq$known'', and in fact would be tested.
On the other hand the r.h.s. corresponds to an input from experimental
data and some theoretical factors likely under control.

For the sake of clarity let us rewrite expression (12) as:
\begin{equation}
\frac{x_j^2\ g(x_j,Q_j^2)^2}{x_i^2\ g(x_i,Q_i^2)^2}\ =\ 
R_{theo}\ {\times}\ R_{exp}
\end{equation}
where
\begin{equation}
R_{theo}(y_{{\Upsilon}i},y_{{\Upsilon}j},p_{Ti},p_{Tj};Q_i^2,Q_j^2)\ =\
\frac{d\hat{\sigma}_{gg}/d\hat{t}_i}{d\hat{\sigma}_{gg}/d\hat{t}_j}
\end{equation}

In the high $p_T$ limit and zero rapidity 
$(y_{{\Upsilon}i}=y_{{\Upsilon}j}=0)$, 
\[ R_{theo}\ {\rightarrow}\ \frac{{\alpha}_s^3(Q_i^2)\ p_{Tj}^4}
{{\alpha}_s^3(Q_j^2)\ p_{Ti}^4}  \]
explicitly showing that ${\alpha}_s(Q^2)$ is entangled in the
gluon density determination. Indeed, 
the determination of $\alpha_s$ from the
analysis of inclusive jet cross sections at hadron colliders
can be viewed as deeply related - in a complementary way - to the 
study of the sensitivity of the hadronic production cross sections 
to different PDF sets \cite{stenzel}. 

Since at high $p_T$ only the $^3S_1^{(8)}$ channel will be
considered (see next Section), the dependence on the NRQCD matrix 
elements does cancel in $R_{theo}$, but there is a dependence on the scales
$Q_i^2$ and $Q_j^2$ which should match the corresponding dependence
in the left hand side. Let us stress that to achieve the
cancellation of NRQCD matrix elements in the quotient, 
only a single channel should contribute dominantly to the cross section.
This fact exludes including subleading 
contributions, preventing us, in principle, from
considering any possible check of next-to-leading
PDFs according to our method, and restricting the test to
leading-order gluon distributions.

On the other hand the experimental input reads as the ratio
\begin{equation}
R_{exp}(p_{Ti},p_{Tj};y_{\Upsilon})= 
\biggl(\frac{p_{Ti}}{p_{Tj}}\biggr)\ {\times}\ 
\biggl(\frac{d^3{\sigma}/dy_{\Upsilon}dy_{jet}dp_{Tj}}
{d^3{\sigma}/dy_{\Upsilon}dy_{jet}dp_{Ti}}\biggr)_{y_{jet}=-y_{\Upsilon}}
\end{equation}
which can be obtained directly from experimental data
\footnote{
Let us note that in the condition $y_{jet}=-y_{\Upsilon}$
imposed to the events - within the uncertainty interval -
could be removed, that is not requiring
a back-to-back topology. This would
amount to extend the study of the ratio (12)
over the region of the plane $x_a,x_b$ 
beyond the diagonal region $x_a=x_b$, thereby allowing 
additional cross-checks of the PDF under scrutiny; however, 
we have not included this study in the present paper.}.

\begin{figure}[htb]
\centerline{\hbox{
\psfig{figure=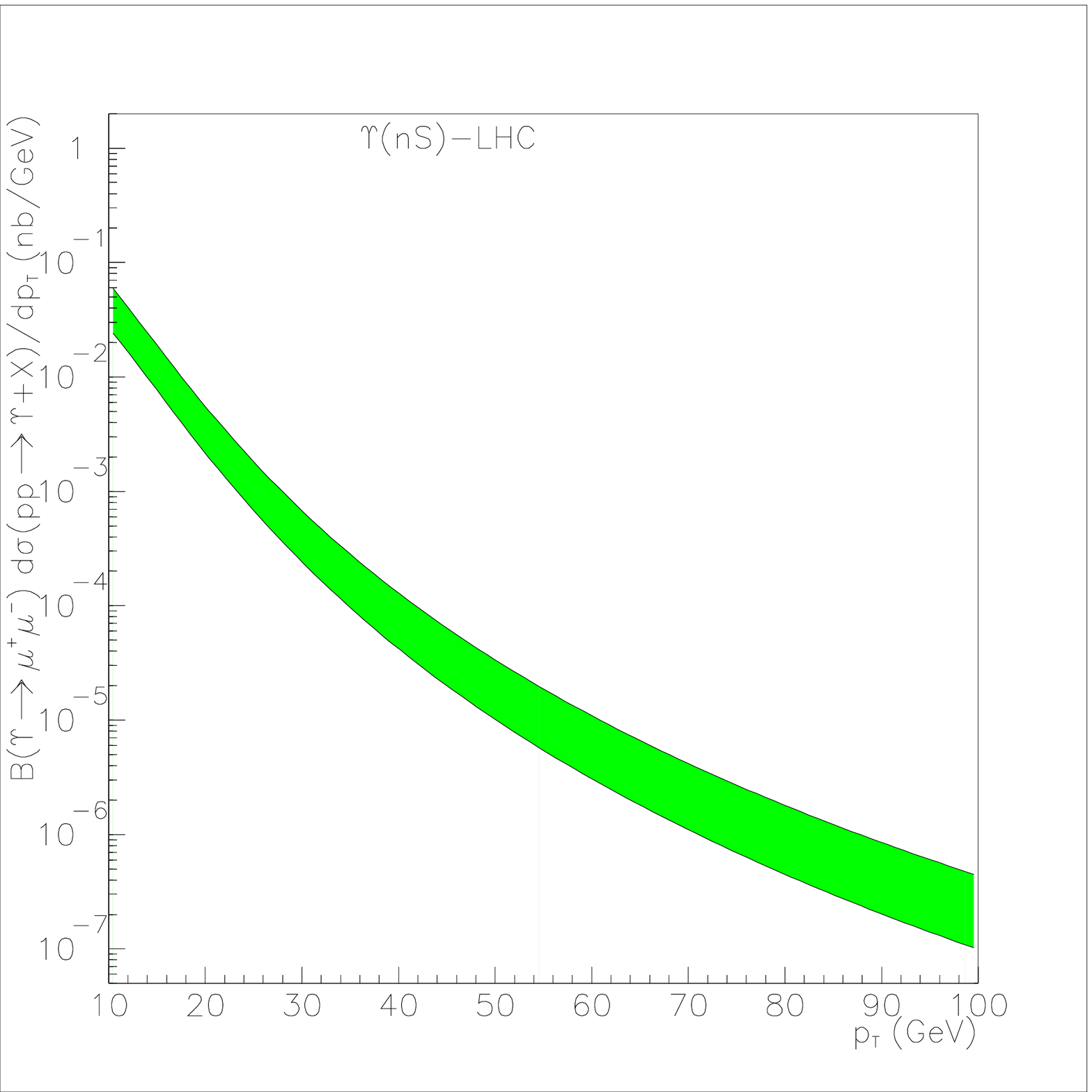,height=7.5cm,width=7.5cm}
\psfig{figure=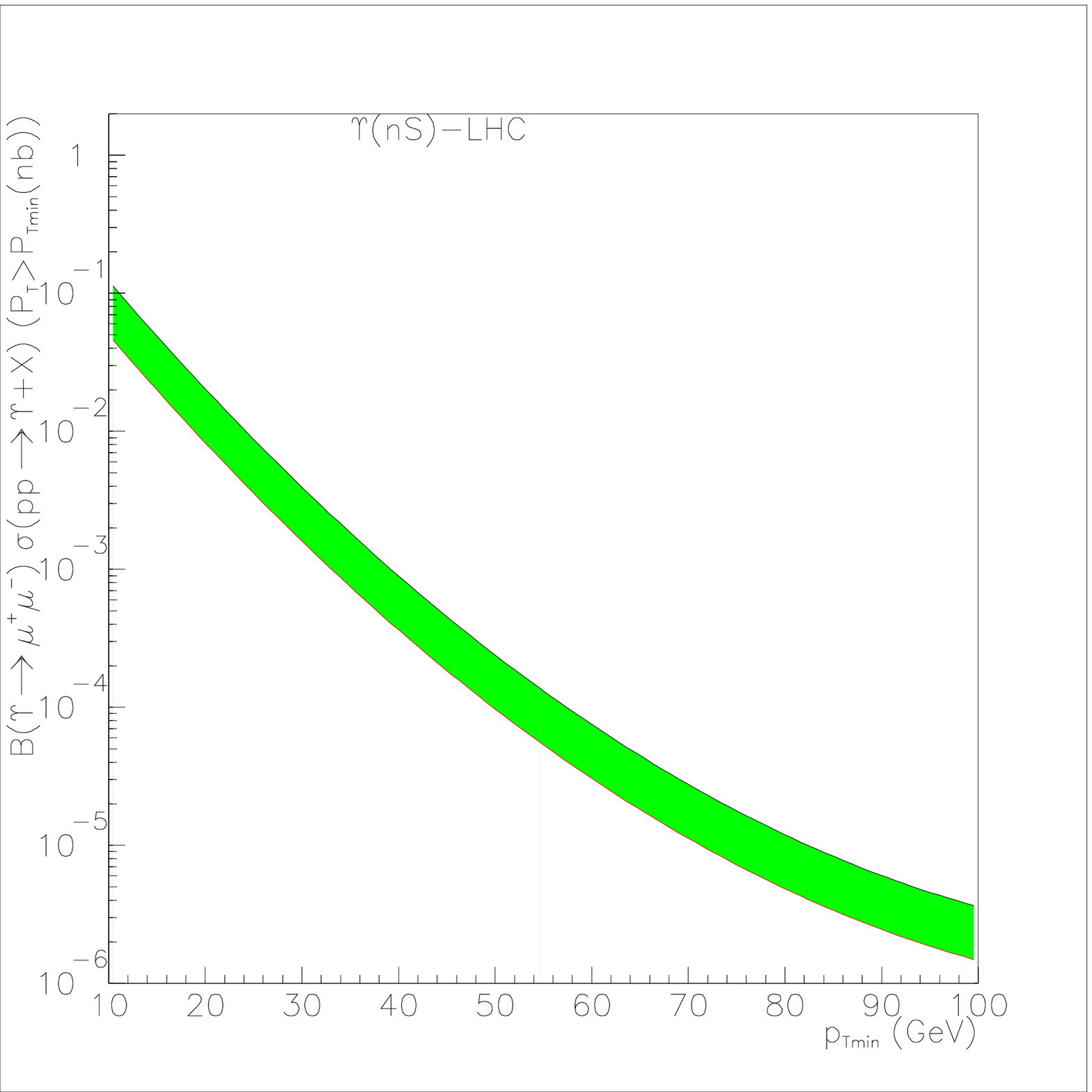,height=7.5cm,width=7.5cm}
}}
\caption{Predicted  $\Upsilon(1S)+\Upsilon(2S)+\Upsilon(3S)$ weighted
contributions to bottomonia inclusive production at the LHC in the
rapidity interval ${\mid}y_{\Upsilon}{\mid}<0.2$ and $p_T>10$ GeV.
Only direct production is considered (no feed-down from
$\chi_{bJ}$ states).
{\em Left panel}: differential cross section; {\em right panel}: integrated
cross section (see Ref.~\cite{mas01}). Let us note that once
the colour-octet MEs have been determined from the Tevatron data, 
using either CTEQ4L or CTEQ5L as the input PDF, the respective predictions 
for the LHC are very similar \cite{prog}.}
\end{figure}

\subsection{Efficiencies, statistics and expected accuracy}

From an experimental point of view, one of the advantages 
of heavy quarkonium physics (bottomonium in particular) at the LHC is
its clean (and self-triggering) signal through the muonic decay
channel: $\Upsilon(nS)\ {\rightarrow}\ \mu^+\mu^-$ allowing a full
kinematic reconstruction. However, it may happen that 
the discrimination among the different 
$\Upsilon(nS)$ states via mass reconstruction could become 
a difficult task at very high $p_T$
because of the uncertainty on the measurement of quite
large momenta of muons. Indeed, above $p_T\ {\simeq}\ 50$ GeV
for each individual muon momenta,  
the uncertainty associated to the reconstruted mass from
two muons makes impossible the discrimination below
the $1$ GeV level \cite{tdr}. Nevertheless, since we are
proposing to study {\em ratios} of cross sections, we can consider 
the overall $\Upsilon(nS)$ direct production, without separating 
the different bottomonia sources - the weighted matrix element cancelling
in the quotient (14) if the mass difference
between the different $\Upsilon$ states is neglected \cite{mas01}. 

In order to assess the foreseen statistics and
efficiency factors, we have employed the PYTHIA  
\footnote{The quarkonium (charmonium) sector 
regarding $2{\rightarrow}2$ production channels
has remained unchanged in PYTHIA from the older  
version 5.7 - used in our earlier studies - to the current one 
6.2 - leading to the same results 
in all cases for high $p_T$.} 
event generator \cite{pythia}
with the colour-octet production mechanism implemented as a 
new code in the Monte Carlo programme (a detailed
account of the implementation, results and discussion can be found in
Ref.~\cite{mas01}). 
For the purpose of illustration we plot in Fig.2 the combined
production rate at $p_T>10$ GeV and ${\mid}y_{\Upsilon}{\mid}<0.2$
for the upper and lower values of the colour-octet
matrix element obtained in \cite{mas01} from previous fits to 
CDF experimental data of Run IB at the Tevatron.

Note that despite the respective muonic
branching ratios are already included in the plot,
actual measurements require
taking into account efficiencies for triggering, reconstruction
and identification of particles and jets from detectors.
In our study we have not performed a full simulation of the
detector effects, although we have
benefitted from different studies focusing on leptons
and jets separately \cite{tdr}. Since we are addressing the feasibility
of this proposal, rather than searching for precise predictions,
this approximation should suffice.

Firstly, the reconstruction of the $\Upsilon(nS)$ mass from the
${\mu}^{+}{\mu}^{-}$ pair implies a reduction factor of about $80\%$ 
by applying the same cuts as for the $J/\psi$ \cite{tdr}, i.e.~a 
$[-3\sigma,+3\sigma]$ symmetric window around the nominal mass. These 
figures do not
include however the muon trigger and identification efficiencies
which altogether roughly amount to $85\%$ for the triggering muon 
and $95\%$ for its partner, respectively.
Therefore, the overall reconstruction factor can be estimated as
\[
\epsilon_{\mu^+\mu^-}\ =\ 0.8\ \times\ 0.85\ \times\ 0.95\ 
\simeq\ 0.65 
\]

Moreover, jet reconstruction efficiency, $\epsilon_{jet}$,  amounts 
on the average to about $75\%$ in the $p_T$ range under 
consideration, using an algorithm with
a jet cone radius ${\Delta}R=\sqrt{{\Delta}{\eta}^2+{\Delta}{\phi}^2}=0.7$
\cite{paco}. 
On the other hand, there is an additional reduction factor, denoted
as $\epsilon_{y}$, due to the requirement of a back-to-back topology
compatible with the constraint given by Eq.(7), which
sets the expected accuracy in the determination of
$x$; choosing $y_{cut}=0.2$ as already mentioned,   
we found (see Ref.~\cite{mas01} for
more details) that $\epsilon_{y}\ {\simeq}\ 35\%$.

In sum, the combined efficiency and
reduction factor to be applied is
\[
\epsilon_{tot}\ =\ \epsilon_{\mu^+\mu^-}\ {\times}\ \epsilon_{jet}\
{\times}\ \epsilon_{y}\ =\ 
0.65\ {\times}\ 0.75\ {\times}\ 0.35\ {\simeq}\ 0.17
\]

Assuming an integrated luminosity of $10$ fb$^{-1}$, corresponding
to one year running ($10^{7}$s) of LHC at $\lq\lq$low'' luminosity
($10^{33}$ cm$^{-2}$s$^{-1}$) and sweeping all the rapidity range
[-2.5,2.5] by steps of 0.4 rapidity units - according to our choice 
$y_{cut}=0.2$ - we find from our simulation 
an expected number of about 10,000 events, corresponding to
the highest Feynman $x$ bin around the value
of $0.014$ considered
in our analysis. Multiplying this number by the
total efficiency factor $\epsilon_{tot}$, we get the estimate:
\[ {\simeq}\ \ 1,700\ events\ per\ year\ run \]

Hence after three years of LHC running at low luminosity
\footnote{We are considering the possibility of extending our proposal to
the high luminosity regime at the LHC coping with
pile-up effects, as well as using the electronic
decay channel of the $\Upsilon$ resonances, too. This would mean to
increase considerably the foreseen statistics allowing
to raise the $x$ range available in our study.}, the 
accumulated statistics 
at the largest $x$ bin would amount to
about $5,000$ events, basically fixing the typical accuracy 
of the order of {\bf ${\simeq}\ 2.5\%$}, which essentially sets the
$\lq\lq$discrimination level'' to be applied to 
different sets of PDFs, as we shall see in the next section.

\begin{figure}[htb]
\centerline{\hbox{
\psfig{figure=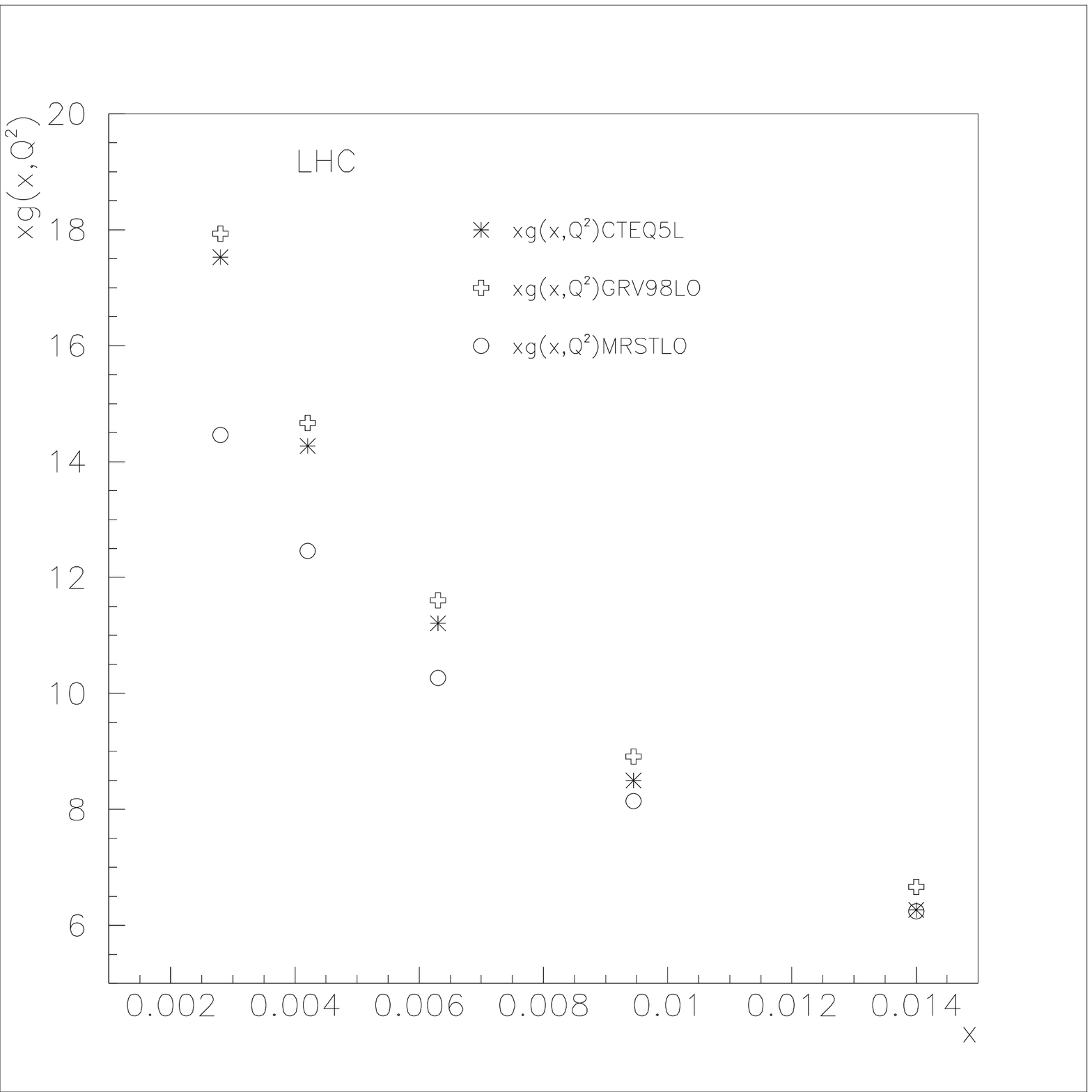,height=7.5cm,width=7.5cm}
\psfig{figure=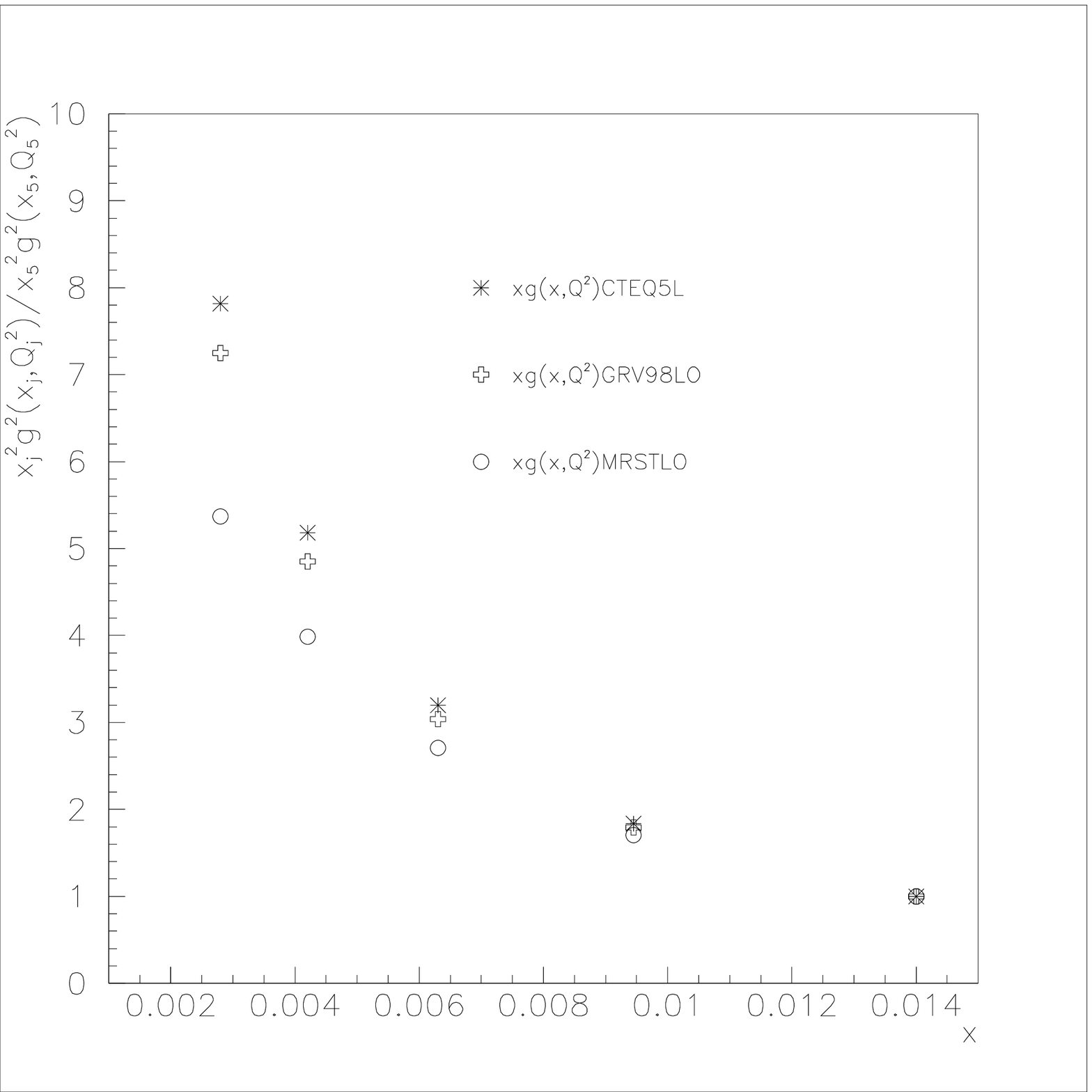,height=7.5cm,width=7.5cm}
}}
\caption{ a) $x_ig(x_i,Q_i^2)$ values for
different leading log (LL) PDFs \cite{durham}. b)
Ratios $x_i^2g^2(x_i,Q_i^2)/x_5^2g^2(x_5,Q_5^2)$, i.e.~values 
of $x_i^2g^2(x_i,Q_i^2)$ normalized to the
rightmost point ($i=5$); $Q_i^2$ represents here an
average value over the whole rapidity
according to Eq.(11) but dependent on $x_i$.}
\end{figure}

\section{Testing the $x$-shape of different PDFs}

As already pointed out, the l.h.s.~of Eq.(12) can be considered as 
an input from a given parameterization set of the gluon distribution
of the proton; 
in Fig.~3.b we show the $x$-shape
for different choices (CTEQ, MRST, GRV)
\cite{durham}. All distributions were
normalized to the rightmost point of Fig.~3.a, i.e.~corresponding 
to $x_5=0.014$. (Although arbitrary in principle, this particular
choice seems quite
appropriate since all PDFs become close each other near this point.)
The points shown in the plot
correspond to different values of the $x$ variable
whose bin widths follow the requirement of Eq.~(7) with
$y_{cut}=0.2$; $Q_i^2$ was selected as the average value of
Eq.~(11) for each $x$ value. For the sake of clarity, let
us stress that the plots in Fig.~3 have a $\lq\lq$general validity'', i.e.
can be obtained from a any PDF parameterization set under
the above-mentioned conditions, without
any reference to a specific experiment;
the applicability to a particular case depends crucially on the 
expected number of events.

In this regard we show in Fig.~4 the maximum difference (in $\%$)
obtained from the curve of Fig.3.b, 
amounting to about $35\%$ at most
over the whole $x$ interval. Therefore 
the foreseen precision level of $2.5\%$ (obtained in the
previous Section) likely should permit
discriminating easily among those LL PDFs at the LHC. Finally, let us
note that, according to our
remark in Section 2, we have not included next-to-leading order PDFs in
our analysis since this would require, for consistency, further subleading
contributions even at high $p_T$, and therefore new NRQCD matrix
elements not completely cancelling in the ratio (14).

On the other hand, it might be possible to perform an overall 
consistency check
on bottomonium production by considering the integrated cross section 
above a lower $p_{Tmin}$ cutoff, say 20 GeV, for
different parton distributions. This is 
in accordance with the
suggestion made in \cite{thorne,martin} of looking at the
uncertainties on physical quantities rather than obtaining errors
associated to the PDFs. In particular, expected differences of
the theoretical calculations on the integrated cross sections 
above $p_{Tmin}=20$ GeV for 
bottomonia hadroproduction at the LHC, using
different parameterization sets, is of order $40\%$
respectively 
\footnote{For a lower $p_{Tmin}$
the difference becomes even larger; for instance, it could reach a factor 2
for $p_T>1$ GeV! However at small/moderate $p_T$ the colour-singlet
mechanism would be largely dominant with quite more parameters 
entering the computation - such as
the wave functions (and derivatives) of bottomonia states,
cascade branching ratios \cite{mas01} - which would make more complex the
physical interpretation of any discrepancy with the experimental result.}.

\begin{figure}[htb]
\centerline{\hbox{
\psfig{figure=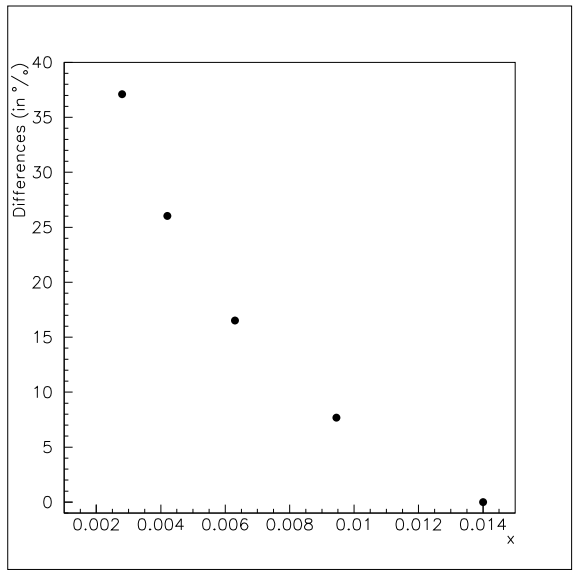,height=7.5cm,width=7.5cm}
}}
\caption { Differences (in $\%$) between current 
$\lq\lq$normalized'' LL PDFs corresponding to
the upper (CTEQ) and lower (MRST) values of Fig.~3.b.
A foreseen precision of ${\simeq}\ 2.5\%$ should suffice to
discriminate among them over the whole $x$ range shown in the plot.}
\end{figure}

\section{Conclusions and last remarks}

With the advent of high energy, high luminosity
proton colliders, 
it is universally recognized the importance of knowing the
uncertainties of the parton distributions to perform
accurate measurements of the Standard Model parameters and
carry out a tantalizing discovery programme on new physics. Keeping this
interest in mind, $\lq\lq$direct'' determinations of the 
gluon density by means of
distinct processes should permit the computation of
systematic and theoretical errors, as well as putting stringent constraints
to global fits where DIS data, however, are still playing a central role.

To this aim and in view of the foreseen rates of bottomonia 
production at relatively large transverse momentum in the LHC \cite{mas01}, 
we propose testing the $x$-shape of the gluon density in protons.
The proposal could be similarly applied to forthcoming 
experimental data from 
the high-luminosity run II at the Tevatron.

If gluon fragmentation is confirmed as the dominant
production bottomonia mechanism at high transverse momentum, the topology 
of events would consist of
an almost isolated lepton pair 
coming from the resonance decay, recoiling against a jet. This represents
basically two advantages: 

\begin{itemize}

\item There is a well reconstructed
direction defined by the lepton pair which can be
assigned to the fragmenting gluon in the final
state of the hard interaction, thereby helping in the
search for the partner jet with a back-to-back topology, 
tagging altogether the event.

\item According to 
the colour-octet production mechanism, the lepton pair should provide,
by means of its measured rapidity and transverse momentum, a
clean kinematic information about the partonic subprocess
itself (i.e.~the $x$-values of the colliding partons). 

\end{itemize}

From inspection of the Figs.~(3-4) and 
the expected discrimination level (${\simeq}\ 2.5\%$)
corresponding to the foreseen statistics collected after three years of
data-taking at the LHC (at $\lq\lq$low'' luminosity), we
concluded that a clear discrimination between different sets of LL PDFs
should be feasible.
In this regard, let us stress that Monte Carlo simulations in
high energy physics have definitely become an
ordinary - as well as necessary - tool nowadays for the 
experimental analysis of data, in particular to disentangle interesting
signals from background. General-purpose event generators 
implementing LL PDFs as default options
(like PYTHIA) are widely used by the 
scientific community involved in the
front-line analysis of experimental data. For instance, cross
section measurements often require Monte Carlo estimates
of the fraction of unobserved events due to detector acceptance
and efficiency limits; an equivalent argument applies to
background estimates. A 
knowledge, as precise as possible, of the dependence of the
theoretical predictions on the choice of a particular PDF
is mandatory. 

Then, we have concluded that our proposal could be useful as
a consistency check of different  
gluon density parameterizations for the proton, of 
interest for other physical
processes involving gluons in the initial state, too. In particular, 
exciting searches for extra dimensions have being recently carried
out at the Tevatron \cite{extra1} and are foreseen
at the LHC \cite{extra2} where effective Planck scales \cite{arkani} 
of the order of the TeV can be tested. Relevant processes in the latter case
are $gg\ {\rightarrow}\ G_{KK}g$, $gq\ {\rightarrow}\ G_{KK}q$, ... 
where $G_{KK}$ denotes a Kaluza-Klein graviton, leading to
a final-state single jet and missing energy. The channel
given by Eqs.(1-2) should also be helpful in the study of the detector response
to a monojet and missing transverse energy, taking into account 
in the latter case the information provided by the recoiling muonic pair.

On the other hand, parton distributions unintegrated over the parton
transverse momentum are increasingly becoming employed in the analysis 
of physical processes initiated by hadrons. Indeed, one of
their advantages is that they correspond to the quantities entering
Feynman diagrams allowing for true kinematics even at leading order.
In our particular case, this would demand to modify the cross section
given in Eq.(3). However, the theoretical situation regarding
heavy quarkonium production is still controversial since there are
important discrepancies between the collinear approximation and
the so-called $k_T$ factorization \cite{teryaev,kraemer}.

A final remark is in order. Although the universality 
of the colour-octet
matrix elements is not definitely well established \cite{kraemer}, one 
could expect that the matrix elements obtained in essentially the same kind of
hadronic processes (i.e.~hadroproduction
at the Tevatron)
should become reliable enough once used at LHC energies, under the same
theoretical inputs (e.g. the bottom mass, choice for the factorization
and fragmentation scales, etc).
Therefore, if a precise normalization of the bottomonia
cross section is
provided (using forthcoming data from high-luminosity Run II
at the Tevatron) the possibility exists of probing different PDF sets 
by means of an unfolding procedure from
heavy quarkonia hadroproduction at the LHC.

\subsection{Acknowledgements} We want to thank
F. Camarena, N. Ellis and S. Tapprogge for interesting discussions.
S.W. acknowledges support from the Deutsche Forschungsgemeinschaft (DFG).

\end{document}